\documentclass[12pt]{article}
\usepackage{html}
\usepackage{epsf}
\usepackage{graphicx}
\usepackage{amssymb}
\usepackage{hyperref}
\begin{document}
\title{MATTERS OF GRAVITY, The newsletter of the APS Division of Gravitational Physics}
\begin{center}
{ \Large {\bf MATTERS OF GRAVITY}}\\ 
\bigskip
\hrule
\medskip
{The newsletter of the Division of Gravitational Physics of the American Physical 
Society}\\
\medskip
{\bf Number 48 \hfill December 2016}
\end{center}
\begin{flushleft}
\tableofcontents
\vfill\eject
\section*{\noindent  Editor\hfill}
David Garfinkle\\
\smallskip
Department of Physics
Oakland University
Rochester, MI 48309\\
Phone: (248) 370-3411\\
Internet: 
\htmladdnormallink{\protect {\tt{garfinkl-at-oakland.edu}}}
{mailto:garfinkl@oakland.edu}\\
WWW: \htmladdnormallink
{\protect {\tt{http://www.oakland.edu/?id=10223\&sid=249\#garfinkle}}}
{http://www.oakland.edu/?id=10223&sid=249\#garfinkle}\\

\section*{\noindent  Associate Editor\hfill}
Greg Comer\\
\smallskip
Department of Physics and Center for Fluids at All Scales,\\
St. Louis University,
St. Louis, MO 63103\\
Phone: (314) 977-8432\\
Internet:
\htmladdnormallink{\protect {\tt{comergl-at-slu.edu}}}
{mailto:comergl@slu.edu}\\
WWW: \htmladdnormallink{\protect {\tt{http://www.slu.edu/colleges/AS/physics/profs/comer.html}}}
{http://www.slu.edu//colleges/AS/physics/profs/comer.html}\\
\bigskip
\hfill ISSN: 1527-3431

\bigskip

DISCLAIMER: The opinions expressed in the articles of this newsletter represent
the views of the authors and are not necessarily the views of APS.
The articles in this newsletter are not peer reviewed.

\begin{rawhtml}
<P>
<BR><HR><P>
\end{rawhtml}
\end{flushleft}
\pagebreak
\section*{Editorial}

The next newsletter is due June 2017.  This and all subsequent
issues will be available on the web at
\htmladdnormallink 
{\protect {\tt {https://files.oakland.edu/users/garfinkl/web/mog/}}}
{https://files.oakland.edu/users/garfinkl/web/mog/} 
All issues before number {\bf 28} are available at
\htmladdnormallink {\protect {\tt {http://www.phys.lsu.edu/mog}}}
{http://www.phys.lsu.edu/mog}

Any ideas for topics
that should be covered by the newsletter should be emailed to me, or 
Greg Comer, or
the relevant correspondent.  Any comments/questions/complaints
about the newsletter should be emailed to me.

A hardcopy of the newsletter is distributed free of charge to the
members of the APS Division of Gravitational Physics upon request (the
default distribution form is via the web) to the secretary of the
Division.  It is considered a lack of etiquette to ask me to mail
you hard copies of the newsletter unless you have exhausted all your
resources to get your copy otherwise.

\hfill David Garfinkle 

\bigbreak

\vspace{-0.8cm}
\parskip=0pt
\section*{Correspondents of Matters of Gravity}
\begin{itemize}
\setlength{\itemsep}{-5pt}
\setlength{\parsep}{0pt}
\item Daniel Holz: Relativistic Astrophysics,
\item Bei-Lok Hu: Quantum Cosmology and Related Topics
\item Veronika Hubeny: String Theory
\item Pedro Marronetti: News from NSF
\item Luis Lehner: Numerical Relativity
\item Jim Isenberg: Mathematical Relativity
\item Katherine Freese: Cosmology
\item Lee Smolin: Quantum Gravity
\item Cliff Will: Confrontation of Theory with Experiment
\item Peter Bender: Space Experiments
\item Jens Gundlach: Laboratory Experiments
\item Warren Johnson: Resonant Mass Gravitational Wave Detectors
\item David Shoemaker: LIGO Project
\item Stan Whitcomb: Gravitational Wave detection
\item Peter Saulson and Jorge Pullin: former editors, correspondents at large.
\end{itemize}
\section*{Division of Gravitational Physics (DGRAV) Authorities}
Chair: Laura Cadonati; Chair-Elect: 
Peter Shawhan; Vice-Chair: Emanuele Berti. 
Secretary-Treasurer: Thomas Baumgarte; Past Chair:  Deirdre Shoemaker;
Members-at-large:
Steven Drasco, Tiffany Summerscales, Duncan Brown, Michele Vallisneri, Kelly Holley-Bockelmann, Leo Stein.
Student Members: Megan Jones, Jessica McIver.
\parskip=10pt

\vfill\eject

\section*{\centerline
{we hear that \dots}}
\addtocontents{toc}{\protect\medskip}
\addtocontents{toc}{\bf DGRAV News:}
\addcontentsline{toc}{subsubsection}{
\it we hear that \dots , by David Garfinkle}
\parskip=3pt
\begin{center}
David Garfinkle, Oakland University
\htmladdnormallink{garfinkl-at-oakland.edu}
{mailto:garfinkl@oakland.edu}
\end{center}

Bob Wald has been awarded the APS Einstein Prize.

Stuart Shapiro has been awarded the APS Bethe Prize.

Martin Rees has been awarded the APS Lilienfeld Prize.

James Bjorken, Sekazi Mtingwa and Anton Piwinski have been awarded the APS Wilson Prize.

Ronald Drever, Kip Thorne, Rainer Weiss and the LIGO team were awarded the Gruber Cosmology Prize and the Special Breakthrough Prize in Fundamental Physics.

Barry Barish, Ronald Drever, Kip Thorne, and Rainer Weiss have been awarded the Smithsonian American Ingenuity Award in the Physical Sciences.

Kip Thorne has been awarded the Tomalla Prize for Extraordinary Contributions to General Relativity and Gravity and the Georges Lemaitre International Prize, and has been included in Time Magazine's 2016 list of the 100 most influential people in the world.

Gabriela Gonz\'alez has been awarded the Jesse W. Beams award  of the Southeastern Section of APS, the Scientist of the year HENAAC award of the Great Minds in STEM organization, the Manne Siegbahn medal of AlbaNova University Center, Stockholm, the Pioneer in Science award of the World Science Festival in New York City, and the ``Brigadier General Juan Bautista Bustos'' award of the C\'ordoba state government. 

Joe Polchinski, Andy Strominger and Cumrun Vafa and have been awarded a Breakthrough Prize in Fundamental Physics.

Frans Pretorius has been awarded a New Horizons in Physics Prize.

The DGRAV bylaws have been approved.

Beverly Berger was elected DGRAV representative to APS.
Gary Horowitz was elected Vice Chair of DGRAV; Geoffrey Lovelace was elected Secretary/Treasurer of DGRAV; Ted Jacobson and Lisa Barsotti were elected members at large of the Executive Committee of DGRAV. Cody Messick was elected Student Representative of DGRAV.

Thomas Baumgarte, Jiri Bicak, Valery Frolov, Paolo Gondolo, Michael Landry, Adrian Lee, Eva Silverstein, Ulrich Sperhake, Robin Stebbins, and Michele Vallisneri have been elected APS Fellows.

Hearty Congratulations!

\vfill\eject
\section*{\centerline
{DGRAV program at the APS}
\centerline{``April'' meeting in Washington D.C.}}
\addtocontents{toc}{\protect\medskip}
\addcontentsline{toc}{subsubsection}{
\it DGRAV program at the APS meeting in Washington D.C. , by David Garfinkle}
\parskip=3pt
\begin{center}
David Garfinkle, Oakland University
\htmladdnormallink{garfinkl-at-oakland.edu}
{mailto:garfinkl@oakland.edu}
\end{center}

We have a very exciting DGRAV related program at the upcoming APS ``April'' meeting January 28-31 in Washington, D.C.  Our Chair-elect, Peter Shawhan did an excellent job of putting together this program.
\vskip0.25truein
One of the plenary sessions is on the topic of black holes:\\

{\bf Black Holes}\\
Laura Cadonati, Einstein's Gift: Stellar Mass Black Holes in the LIGO Era\\
Chung-Pei Ma, The Most Massive Black Holes in the Local Universe\\
Andrew Strominger, The Black Hole Information Paradox, Revisited\\

The invited sessions sponsored by DGRAV are as follows:\\

{\bf Electromagnetic Signatures of Neutron Star Mergers}\\
(joint with DAP and DNP)\\
Rebecca Surman, Nucleosynthesis and Neutrino Physics in Compact Object Mergers\\
Andreas Bauswein, Dynamics of Compact Object Mergers\\
Luke Roberts, Radioactive Powered Transients from Compact Object Mergers\\

{\bf Extremes of Gravity: from Weak to Strong}\\
Lydia Bieri, Insights into the Gravitational Wave Memory Effect\\
Samuel Gralla, Growth of Perturbations Near a Rapidly Spinning Black Hole\\
James Isenberg, Strong Cosmic Censorship\\

{\bf Einstein Prize Talk and Advanced LIGO Search Results}\\
Robert Wald, Black Holes, Thermodynamics, and Quantum Theory\\
Jocelyn Read, Searches for all Types of Binary Mergers in the First\\
 Advanced LIGO Observing Run\\
Patrick Meyers, Searching for the Stochastic Gravitational Wave Background\\
 in LIGO's First Observing Run\\

{\bf Observers and Entanglement in Spacetime}\\
William Donnelly, Implications of Diffeomorphism Invariance for Observables in Gravity\\
Eugenio Bianchi, Entanglement and the Architecture of Spacetime\\

{\bf Numerical GR Simulations of Core Collapse Supernovae}\\
(joint with DCOMP)\\
Phillipp M\"osta, (Extreme) Core-collapse Supernova Simulations\\
Bernhard Mueller, Core-collapse Supernova Simulations\\
Anthony Mezzacappa, Modelling Core Collapse Supernovae\\

{\bf Gravitational Wave Observations with Current and Future Facilities}\\
Joseph Giaime, Ground-based Gravitational-wave Observatories\\
Laura Sampson, Pulsar Timing Arrays: Closing in on Low-frequency Gravitational Waves\\
Guido Mueller, The Path to a Gravitational-wave Detector in Space\\

{\bf Progenitors of Merging Binary Black Holes}\\
(joint with DAP)\\
Selma de Mink, Classical Binary Star Evolution Leading to a Binary Black Hole\\
Nicholas Stone, Dynamical Formation and Merger of Binary Black Holes\\
Carl Rodriguez, Binary Black Holes from Dense Star Clusters\\

{\bf Numerical Relativity Simulations of Compact Binary Mergers}\\
(joint with DCOMP)\\
Geoffrey Lovelace, Simulations of Binary Black Hole Mergers\\
Matthew Duez, The Exotic Remnants of Compact Object Binary Mergers\\
Kenta Kiuchi, Simulations of Binary Neutron Star Mergers\\

The focus sessions sponsored by DGRAV are as follows:
\vskip0.25truein
{\bf
Testing General Relativity with Black Hole Observations\\
Electromagnetic Follow-up of Gravitational Wave Candidate Events\\
Loop Quantum Gravity and Cosmology\\
}

The contributed sessions sponsored by DGRAV are as follows:
\vskip0.25truein
{\bf
Tests of General Relativity and Alternatives: Theory Meets Experiment\\
Gravitational-wave Detection in Space: from LISA Pathfinder to LISA\\
Gravitational-wave Astrophysics: Neutron Stars and Core Collapse Supernovae\\
Numerical Relativity Codes and Methods\\
LIGO/Virgo and Holometer Searches for Continuous-wave,\\
Burst and Supernova Signals\\
Getting a Handle on the Population of Merging Binaries\\
Orbits, Spins and Tides: Analytic and Numerical Methods\\
New Directions in Gravity\\
Numerical Relativity Solutions for Black Hole Binaries\\
Primordial Black Holes, Spacetime Structure and Thermodynamics\\
Quantum Aspects of Gravitation\\
Black Holes and their Environments\\
Mathematical Aspects of General Relativity\\
Gravitational Wave Source Modeling\\
Numerical Relativity Simulations of Neutron Star Binaries\\
 and Gravitational Collapse\\
Gravitational Wave Data Analysis Techniques\\
Achieving Optimal Sensitivity with LIGO\\
 and other High-frequency Gravitational-wave Detectors\\
}

\vfill\eject

\section*{\centerline
{Remembering John Stewart}}
\addtocontents{toc}{\protect\medskip}
\addtocontents{toc}{\bf Obituary:}
\addcontentsline{toc}{subsubsection}{
\it  Remembering John Stewart, by Paul Shellard}
\parskip=3pt
\begin{center}
Paul Shellard, University of Cambridge 
\htmladdnormallink{E.P.S.Shellard-at-damtp.cam.ac.uk}
{mailto:E.P.S.Shellard@damtp.cam.ac.uk}
\end{center}

It is with sadness that we report on the passing of John Stewart, Emeritus Reader of Gravitational Physics and Fellow of King's College.

John Stewart was born in Pinner, west London. He was educated at Latymer Upper School and studied at Jesus College, Cambridge, receiving his BA in 1965. He became a student of Dennis Sciama and then George Ellis, graduating with his PhD from the Department of Applied Mathematics and Theoretical Physics (DAMTP) in 1969. From 1968-71 he was a Junior Research Fellow at Sidney Sussex College, moving then to Munich to take up a postdoctoral fellowship at the Max Planck Institute for Physics and Astrophysics. He returned to Cambridge and became a DAMTP Lecturer in 1976. He was promoted to a Senior Lectureship in 2000 and to a Readership in Gravitational Physics in 2003.  

For forty years, John was a member and a pillar of the Relativity and Gravitation (GR) group in DAMTP, contributing to its teaching and research activities. Throughout his career, he was universally admired as a superb lecturer by successive generations of students, due to his careful systematic style and the clarity of his explanations, as well as his sense of humour. He pioneered research in relativistic kinetic theory, cosmological perturbation theory and numerical relativity over many years in Cambridge. A number of his former graduate students continue to pursue research in relativity in academic posts around the world, including Luke Drury (Dublin IAS), Ruth Gregory (Durham), Ian Hawke (Southampton) and Oliver Rinne (Albert Einstein Institute). Research activity in numerical relativity remains vibrant in DAMTP and, of course, John was very interested in the recent LIGO announcement of the direct detection of gravitational waves.

John Stewart was also a Life Fellow of King's College, having become a Senior Research Fellow there in 1975 and then a Fellow and College Lecturer in 1978. Professor Michael Proctor, Provost of King's, writes:

``John retired as Reader in Gravitational Physics in 2010, but continued to be active in the College and Department until the week before his death. His work on numerical relativity was much admired, but he will be best known to generations of mathematics students at King's and elsewhere as a patient, careful and meticulous teacher, able to bring illumination to the most difficult of subjects. He took a keen and thoughtful interest in the College's finances and served on the Investment and Finance Committees and as Inspector of Accounts. He was also very fond of the musical life of the College and was a regular attender at concerts. His expertise will be much missed. He leaves a widow, Mary, a Fellow of Robinson College.''

John began his research career studying relativistic kinetic theory and hydrodynamics, culminating in his 1971 book on ``Non-equilibrium relativistic kinetic theory'' (in the Springer Lecture Notes in Physics series). In 1967 with Dennis Sciama, he published his first paper in Nature on the dipole temperature variations expected in the cosmic microwave background due to the peculiar velocity of the Sun, thus defining an absolute isotropic frame of reference for our Universe. In the same year, with George Ellis, he found a family of perfect fluid and magnetic exact anisotropic cosmological solutions of Einstein's equations that have been important in the study and classification of spatially homogeneous universes.

From 1968-1971, John studied the role of collisionless particles and non-equilibrium behaviours in anisotropic universes and the differences between the behaviour of collisionless particles and that of simple fluid viscosities. This was needed for a detailed analysis of Misner's chaotic cosmology programme and other possibilities for singularity avoidance. John also pioneered the use of phase plane methods in studying the qualitative dynamics of homogeneous cosmologies.  From 1973 onwards, his interest grew in black hole spacetimes, accretion disks and turbulence, and in relativistic perturbation theory.  A particular highlight was his 1974 paper with Martin Walker that developed the mathematical principles underlying gauge-invariant perturbation theory, including the scalar-vector-tensor decomposition, and which has become the standard methodology in the field.

In 1979 with Werner Israel, John developed a comprehensive theory of causal thermodynamics to handle close to equilibrium behaviour in general relativity, including bulk and shear viscosities, heat conduction and shock wave formation. This work, with over 700 citations, has found recent new applications in string theory with the fluid/gravity correspondence.

In the early 1980s, John began a programme of work in numerical relativity starting with papers with Helmut Friedrich studying the characteristic initial value problem. He subsequently investigated gravitational shocks and gravitational waves, and solved gravitational collapse problems in spherically and axisymmetric contexts. A fascinating byproduct was early numerical work on the consequences of inflationary cosmology. In his 1982 paper with Stephen Hawking and Ian Moss on `Bubble collisions in the very early universe', John showed that slow thermalization or black hole formation could strongly constrain the viability of inflation models. Later with Gary Gibbons and Stephen Hawking, he proposed a natural measure on Friedmann cosmologies to study the likelihood of inflation.

In 1991, John published his CUP book on ``Advanced General Relativity'' in the Monographs in Mathematical Physics series.   From 1993 onwards, with David Salopek and his graduate students, he developed Hamilton-Jacobi theory for studying gravitational perturbations in stochastic inflation and quantum cosmology, as well as structure formation in the late universe. Together with graduate students and collaborators, he continued an active research programme in numerical relativity until his retirement, including work on critical phenomena during black hole formation and the development of gauge formulations and augmented systems suitable for numerical evolution.

Despite his retirement in September 2010, John regularly came to DAMTP to have coffee and work in his office. During this time he wrote a CUP book on the Python programming language, ``Python for Scientists'', and he was nearing completion of a much larger and extended version with further scientific applications; he was working on this up to a few days before being taken ill.

John will be very much missed, both in DAMTP and at King's.

\vfill\eject

\section*{\centerline
{The Seventh Meeting on CPT and Lorentz Symmetry}}
\addtocontents{toc}{\protect\medskip}
\addtocontents{toc}{\bf Conference reports:}
\addtocontents{toc}{\protect\medskip}
\addcontentsline{toc}{subsubsection}{
\it CPT and Lorentz Symmetry, 
by Jay Tasson}
\parskip=3pt
\begin{center}
Jay Tasson, Carleton College 
\htmladdnormallink{jtasson-at-carleton.edu}
{mailto:jtasson@carleton.edu}
\end{center}

The seventh meeting on CPT and Lorentz symmetry
was held at Indiana University during the week of June 20th, 2016.
The meeting focused on tests of these fundamental symmetries
and the related theoretical issues across all areas of physics,
including a major component on local Lorentz invariance
in gravitational phenomena.
Approximately 120 people attended the meeting,
over 60 invited and contributed talks were given,
and nearly 30 posters were presented.
Beyond the official program,
many productive and friendly conversations occurred
during the many lunches and coffee breaks.
Written versions of the presentations will be published in the volume
{\it Proceedings of the Seventh Meeting on
CPT and Lorentz Symmetry} \cite{proc},
most of which have also appeared on the arXiv.
The remainder of this report summarizes
some of the key ideas discussed at the meeting.

As an ingredient in the Einstein Equivalence Principle,
local Lorentz invariance
lies at the foundation of General Relativity (GR).
Violations of local Lorentz invariance
may arise in an underlying theory of quantum gravity.
Hence tests of this symmetry
provide a fundamental test of GR
and may provide clues about the nature of an underlying theory
that may merge GR and quantum physics.

One popular approach to searching for Lorentz violation in nature
is via a systematic search within a framework containing known physics
along with all possible Lorentz-violating effects.
An effective field-theory construction known as the gravitational
Standard-Model Extension (SME) provides
such a comprehensive framework \cite{sme},
and many of the talks at the meeting presented
theoretical conclusions and the results of experimental and observational searches
in this context.
Foundational ideas about the systematic search for Lorentz violation
were presented by organizer Alan Kosteleck\'y \cite{ak},
and summarized by many speakers at the meeting.
The status of constraints on Lorentz-violating terms in the effective field-theory expansion
and the rate of progress in the field was presented
in a talk on the annually updated publication
{\it Data Tables for Lorentz and CPT Violation} \cite{data}.

Following the much-anticipated direct observation of gravitational waves,
a number of experimental and phenomenological talks addressed
the implications of gravitational waves
and the LIGO/VIRGO detectors in searches for Lorentz violation.
After a discussion of the basic features of the initial detection,
uses of the observed waveform
in constraining modified gravity,
including Lorentz-violating modifications,
were presented \cite{ny}.
The general linearized effective field theory for gravity
was recently developed and initial applications
to the propagation of gravitational waves
yielded constraints on Lorentz-violating terms
through the absence of evidence for birefringence in the initial 
LIGO events.
In addition to discussion of this action-based approach \cite{mm},
a related search effort based on an expansion
at the level of the dispersion relation
was also presented \cite{rt}.
The lack of evidence for \v Cerenkov gravitational radiation
by high-energy cosmic rays
offers another means of constraining Lorentz-violating terms
associated with gravitational radiation \cite{jt}.
Finally,
as a very different application of LIGO data,
LIGO's sensitivity to low-frequency changes 
in the effective index of refraction of light
was discussed \cite{am}.
This observation led to about 4 orders of magnitude improvement
in constraints on certain Lorentz-violating terms
in the photon sector.

Experimental efforts on a very different scale,
laboratory short-range gravity experiments,
have also made recent and impressive progress in obtaining sensitivity
to Lorentz-violating effects in the gravity sector that are complementary to
those explored using gravitational radiation.
Improvements of about an order of magnitude
over prior work were announced at the meeting.
Several relevant experiments \cite{rd,jl,pl,cgs}
were discussed along with the associated theoretical work \cite{qb,rx}.
In a proposal for very different type of laboratory test
of gravity-sector effects,
it was pointed out that Lorentz-violating effects that mimic gravitomagnetism
may be detectable in the developing GINGER ring-laser experiment \cite{ns}.

Solar-system tests have a long history in testing GR,
and new results in this area were also prominent at the meeting.
Anisotropies in the gravitational field
of solar system bodies are a natural consequence of Lorentz violation in the gravity sector,
and solar-system observations form a laboratory for seeking these effects.
Hence we heard a talk on the modeling and analysis 
of lunar laser ranging data \cite{br},
and in another talk,
new constraints on Lorentz violation
from lunar laser ranging,
very long baseline interferometry, and planetary motion
were presented \cite{cpl}.
Anisotropy in the Earth's gravitational field induced by Lorentz violation
can also be sought in the laboratory.
Tests of this type considered at the meeting
included superconducting gravimeter experiments,
for which preliminary Lorentz-violation sensitivities were announced \cite{jt},
and
atom interferometer tests \cite{cg,hm}.

The notion of a comprehensive framework for describing Lorentz violation
leads naturally to effects that depend on the composition
of the matter used in gravitational experiments,
and hence to violations of the equivalence principle.
The point is that in the matter sector,
terms arise involving couplings between matter,
Lorentz violation, and gravity.
The full generality of the framework allows different couplings
to Lorentz violation
for different types of matter.
Hence in addition to pure-gravity effects,
Lorentz violation in the matter sector
is interesting for certain gravitational tests,
including many of the same ones
relevant for the pure-gravity sector discussed above.
Additionally,
experiments testing universality of free fall
are sensitive to these effects.
After some discussion of the basic theory
for these types of tests \cite{jt},
some up-coming experiments were considered,
including the Microscope space mission now in progress,
for which a Lorentz-violation analysis is planned \cite{cg}.
Weak Equivalence Principle experiments with antimatter and higher-generation matter
may attain special sensitivities in this context.
Experiments involving antihydrogen \cite{pc,md},
gravitational tests with muons \cite{dk,kk},
and indirect tests with antimatter \cite{tk}
were all presented.

On the theory side,
a number of additional ideas about Lorentz violation in gravity
were developed.
Lorentz violation is typically not compatible with conventional
Riemannian geometry \cite{sme},
and models illustrating this feature
were presented \cite{rb}.
These considerations
have triggered interest in the geometric structure
of Lorentz-violating theories more generally,
and hence Finsler spaces discovered through exploration
of the SME were also among the theory topics discussed \cite{ak,jf}.
Gravitational radiation came up on the theory side as well
through consideration of gravitational wave generation
in various Lorentz-violating models \cite{ny}.
Much work on Lorentz violation in gravity has been done 
in the linearized-gravity limit.
Hence two talks developing nonlinear gravity
were of interest,
one on initial considerations
for
Lorentz-violating effects in the effective field-theory framework 
of the SME \cite{qb},
and another that more abstractly explored the development 
of nonlinear theories of gravity with Lorentz violation
via the boostrap approach \cite{ms}.
Returning to the linearized gravity limit,
a term exists in the general effective field-theory framework
that does not generate physical effects
in phenomenological work to date.
The reason for this lack of phenomenological effects remains mysterious
and is known as the `t-puzzle',
but an update on the situation was provided \cite{yb}.
In the matter sector,
initial work on extending efforts
to include Lorentz-violation effects
with arbitrary spacetime dependence
was presented \cite{cl}.

At this,
the seventh meeting in this triennial series
and the fifth that I attended,
I was once again struck by the ever increasing depth and breadth of the field.
We saw expansion of theory and phenomenology into gravitational wave physics,
and experimental measurements reaching once-unthinkable levels of sensitivity
were discussed,
often probing well beyond Planck-scale sensitivity.
I left the meeting inspired with new ideas for
testing known physics with ever increasing sensitivity
and a heightened enthusiasm to continue the search for new physics
at the quantum-gravity scale.

\end{document}